\documentclass[aps,twocolumn,nofootinbib]{revtex4}
\usepackage[latin1]{inputenc}
\usepackage{epsfig}
\usepackage{mathrsfs}
\newcommand{\beq}{\begin{equation}}
\newcommand{\eeq}{\end{equation}}
\newcommand{\la}{\langle}
\newcommand{\ra}{\rangle}

\begin{document}

\title{Classical stochastic representation of
quantum mechanics}
\author{Mário J. de Oliveira}
\affiliation{Universidade de São Paulo,
Instituto de Física, Rua do Matão, 1371,
05508-090 São Paulo, SP, Brazil}

\begin{abstract}

We show that the dynamics of a quantum system can be
represented by the dynamics of an underlying classical
systems obeying the Hamilton equations of motion.
This is achieved  by transforming the phase space of
dimension $2n$ into a Hilbert space of dimension $n$
which is obtained by a peculiar canonical transformation
that changes a pair of real canonical variables into a
pair of complex canonical variables which are complex
conjugate of each other. The probabilistic character
of quantum mechanics is devised by treating the wave
function as a stochastic variable. The dynamics of
the underlying system is chosen so as to preserve
the norm of the state vector.

\end{abstract}

\maketitle


The earliest formulations of quantum mechanics were   
given by Schrödinger, who introduced the quantum wave
equation that bears his name, and by Heisenberg who
introduced the quantum matrix mechanics. These two
formulations were shown to be equivalent and, together
with other formulations, are understood as different
representations of the same theory. The standard
formulation of quantum mechanics considers that the
quantum state is a vector with complex components
belonging to the Hilbert vector space. The time
evolution of the quantum state is given by a unitary
transformation whose generator is the Hamiltonian
operator acting on the Hilbert space. This type
of evolution guarantees that the norm of the
state vector is preserved for all times.

Quantum mechanics 
\cite{landau1958,merzbacher1961,messiah1961,sakurai1967,
sakurai1994,griffiths1995,piza2002,griffiths2002}
as a science of motion differs fundamentally from
classical mechanics
\cite{lanczos1949,goldstein1950,landau1960,arnold1978}.
For instance, the mathematical objects corresponding
to real physical quantities such as position and
momentum are very distinct in the two theories.
In quantum mechanics they are operators acting on a
Hilbert space and the possible outcomes of an observable
are the eigenvalues of the corresponding operator. 
In fact classical and quantum mechanics are conflicting
scientific theories of the same real phenomena and just
one of them could give the correct predictions. It is 
admitted indeed that classical mechanics does not
correctly describes nature at small scales. At large
scales quantum mechanics is reduced to classical
mechanics and both predict the same results.

The question we address here is not whether the
science of quantum mechanics is equivalent to
the science of classical mechanics, as they are
not. The question we address is whether the abstract
framework of quantum mechanics can in some sense be
equivalent to the abstract framework of classical
mechanics. We answer this question positively by
showing that the dynamics of a state vector
belonging to a Hilbert space of dimension $n$
is equivalent to the dynamics of a classical
system with $n$ degrees of freedom. This classical
system we call the {\it underlying} system to
avoid confusion with a real system described by
classical mechanics. 
The wave equation is then understood as 
related to a pair of classical canonical variables,
its real and imaginary parts being proportional 
to the coordinate and momentum, respectively.
The underlying system cannot be any classical
system, but only those whose motion preserves
the norm of the complex wave function.

The idea of expressing classical mechanics in a
Hilbert space that we use here was considered by
Koopman \cite{koopman1931} who showed that
canonical transformation are equivalent to unitary
transformation if the state functions in phase space
are square integrable \cite{holland1993}. This result
was also used by von Neumann to formulate classical
mechanics as an operational theory
\cite{vonneumman1932}.

Quantum mechanics has a probabilistic character that
is particularly manifest in the standard interpretation
of quantum mechanics according to which the square of
the absolute value of the wave function is a probability.
Here, the probabilistic character of thermodynamics is
devised by considering that the wave function is a
stochastic variable, that is, a time dependent random
variable. Accordingly, the wave vector in the Hilbert
follows a stochastic trajectory. In this sense, the
present stochastic representation is in accordance
with the consistent history interpretation of quantum
mechanics \cite{griffiths2002}.

Let us consider the representation of classical mechanics
by the Hamilton equations of motion. In this representation,
a state is defined as a vector of the phase space
spanned by the canonical variables. The dimension of
the vector phase space equals $2n$ where $n$ is the
number of degrees of freedom, which is the number
of pairs of canonically conjugate variables. The
canonical Hamilton equations of motion are given by 
\beq
\frac{dq_k}{dt} = \frac{\partial{\cal H}}
{\partial p_k}, \qquad \frac{dp_k}{dt}
= - \frac{\partial{\cal H}}{\partial q_k},
\eeq 
where ${\cal H}$ is the Hamiltonian function 
and $(q_k,p_k)$ denotes one of the $n$ pairs of
canonically conjugate variables.

The pairwise formulation of the canonical equations
of motion allows a {\it peculiar} transformation 
\cite{lanczos1949} of the pair of real canonical
variables $(q_k,p_k)$ to a pair of complex
canonical variables $(z_k,z_k^*)$. This peculiar
transformation is accomplished by  
$z_k = \alpha_k q_k + i \beta_k p_k$
where  $\alpha_k$ and $\beta_k$ are real constants
such that $\alpha_k\beta_k=1/2\mu$, and $\mu$
is some constant with the physical dimension of
coordinate$\times$momentum. This transformation
guarantees that the pair $(z_k,z_k^*)$ is a pair
of canonically conjugate variables. In terms of
the new variables the equations of motion become
\beq
i \mu \frac{dz_k}{dt}
= \frac{\partial{\cal H}}{\partial z_k^*},
\qquad
i\mu \frac{dz_k^*}{dt}
= - \frac{\partial{\cal H}}{\partial z_k},
\label{13}
\eeq 
where $z_k$ and $z_k^*$ are dimensionless and
treated as independent
variables, and ${\cal H}$ is a real function of
the set of variables $\{z_k\}$ and $\{z_k^*\}$.
The Hamilton equations can also be written in
terms of Poisson brackets
\beq
i\mu\frac{d z_k}{dt} = \{z_k,{\cal H}\},
\qquad
i\mu\frac{d z_k^*}{dt} = \{z_k^*,{\cal H}\}.
\label{11}
\eeq
The Poisson brackets between two state functions 
$\cal A$ and $\cal B$ are defined by
\beq
\{{\cal A},{\cal B}\} =
\sum_j\left(\frac{\partial{\cal A}}{\partial z_j}
\frac{\partial{\cal B}}{\partial z_j^*}
- \frac{\partial{\cal B}}{\partial z_j}
\frac{\partial{\cal A}}{\partial z_j^*}
\right),
\eeq
and we remark that $\{z_j,z_k^*\}=\delta_{jk}$.

The time evolution of a state function ${\cal A}$,
that is, as a function of the set of variables
$\{z_j\}$ and $\{z_k^*\}$, is given in terms of
the Poisson brackets by
\beq
i\mu\frac{d{\cal A}}{dt} = \{{\cal A},{\cal H}\}.
\label{7}
\eeq
which follows from (\ref{11}).

As the two equations of motion for $z_k$ and
$z_k^*$ are the complex conjugate of each other,
we may consider them to be just one equation in
complex variables. Thus we are representing the 
motion of a classical system as a trajectory in
a vector space with $n$ dimensions with complex
components $z_k$, which defines a Hilbert vector
space.

We assume the Hamiltonian ${\cal H}$ to be a
bilinear function in the complex variables,
\beq
{\cal H} = \sum_{jk}H_{jk} z_j^* z_k,
\label{4}
\eeq
where $H_{jk}$ are understood as the elements of a
matrix $H$, which is Hermitian because ${\cal H}$
is real. The norm ${\cal N}$ of a state $\{z_k\}$
is defined by 
\beq
{\cal N} = \sum_j z_j^* z_j,
\label{8}
\eeq
and we see that it is a constant of the motion 
since it commutes in the Poisson sense with the
Hamiltonian, $\{{\cal N},{\cal H}\} = 0$. 
Therefore we may set ${\cal N}$ equal to a
constant which we choose to be 1.

If we replace the expression of ${\cal H}$ given
by the equation (\ref{4}) into the equation of
motion (\ref{11}) we reach the equation
\beq
i\mu\frac{dz_j}{dt} = \sum_k H_{jk}z_k,
\label{9}
\eeq
The variables $z_k$ are understood as the components
of a state vector $\psi$ of the Hilbert, that is,
\beq
\psi = \sum_j z_j \phi_j.
\label{10}
\eeq
where the vectors $\{\phi_j\}$ form a complete
basis of the Hilbert space.
Defining the operator $\hat{H}$ by 
\beq
\hat{H} \phi_k =  \sum_j H_{jk} \phi_j,
\eeq
the equation (\ref{9}) acquires the form
\beq
i\mu\frac{d}{dt} \psi= \hat{H}\psi,
\eeq
which is the Schrödinger equation if we set
the constant $\mu$ equal to the Planck constant,
\beq
\mu = \hbar.
\eeq

In accordance with the postulates of quantum mechanics,
the possible outcomes of an observable $\mathscr{A}$
are the eigenvalues of a matrix $A$. If the system
is in a state $\psi$, given by (\ref{10}), 
the quantum average of this observable is given by 
\beq
{\cal A} = \sum_{jk} A_{jk} z_j^*z_k,
\eeq
where $A_{jk}$ are the elements of a matrix $A$
whose eigenvalues are the possible outcomes of the
observable $\mathscr{A}$. We interpret ${\cal A}$
as a state function related to the underlying 
classical system. 

In the following we change the equations of motion
for the purpose of treating the dynamic variables
as stochastic variables, which we now denote by
$x_k$. This is accomplished by adding a white
noise to the equation (\ref{13}). We choose a
noise that changes the phase $\theta_j$ of
$x_j=r_je^{i\theta_j}$ but not its absolute
value $r_j$.  This is accomplished by writing
equation (\ref{13}) in the polar form
\beq
\frac{d\theta_j}{dt} = \frac1{2\mu r_j}
\frac{\partial{\cal H}}{\partial r_j} + \zeta_j,
\label{15a}
\eeq
\beq
\frac{dr_j}{dt} = - \frac1{2\mu r_j}
\frac{\partial{\cal H}}{\partial\theta_j},
\label{15b}
\eeq
where $\zeta_j$ is a stochastic variables with zero
mean, and the Hamiltonian function is given by
\beq
{\cal H} = \sum_{jk} H_{jk} x_j^* x_k.
\eeq

From the stochastic equation one can obtain the equation
that governs the time evolution of the probability
distribution \cite{vankampen1981,tome2015}. In the
present case the probability distribution, which we
denote by $\mathscr{P}(x,x^*)$, is defined on the
Hilbert space.  The probability distribution obeys
the Fokker-Planck equation
\beq
\frac{\partial\mathscr{P}}{\partial t} = 
\frac1{i\mu}\{\mathscr{P},{\cal H}\} 
+\frac12 \sum_{jk} \gamma_{jk}
\frac{\partial^2\mathscr{P}}
{\partial\theta_j\partial\theta_k},
\label{22}
\eeq
where $\gamma_{jk}=\gamma_{kj}\geq0$.

As the noise $\zeta_i$ does not change $x_j^*x_j$,
it will not change the norm
\beq
{\cal N} = \sum_j x_j^* x_j,
\label{33}
\eeq
and taking into account that $\{{\cal N},{\cal H}\}=0$,
we conclude that ${\cal N}$ is strictly constant along
a trajectory in the Hilbert space, despite the fact that
the trajectory is stochastic. This result allows us to
choose the norm to be equal to 1. We will also choose
the constants $\gamma_{jk}$  to be all equal so that the
noise will not change the phase of $x_jx_k^*$.

The solution of the Fokker-Planck equation (\ref{22})
is a multivariate Gaussian distribution in the variables
$x_j$ and $x_k^*$. Therefore, to construct
$\mathscr{P}(x,x^*)$, it suffices to determined the
averages $\la x_j\ra$ and the covariances
$\rho_{jk}= \la x_j x_k^*\ra$. From equation (\ref{22})
we reach the equations 
\beq
\frac{d}{dt}  \la x_j\ra
= \frac1{i\mu}\sum_k H_{jk} \la x_k\ra
- \frac{\gamma}2 \la x_j\ra,
\label{20}
\eeq
\beq
i\mu \frac{d}{dt}\rho_{jk} =  \sum_\ell
(H_{j\ell} \rho_{\ell k} - \rho_{j\ell} H_{\ell k}),
\label{21}
\eeq
and we remark that there is no term corresponding
to the noise in the last equation due to our
choice of the same value of $\gamma_{jk}=\gamma$.
Taking into account that the norm (\ref{33}) 
equals the unity then
\beq
\sum_j \rho_{jj} = 1.
\label{33a}
\eeq

Defining $\rho$ as the matrix with elements $\rho_{jk}$,
the last equation gives ${\rm Tr}\rho=1$, and
the equation (\ref{21}) acquires the form
\beq
i\mu \frac{d\rho}{dt} = [H,\rho].
\label{62}
\eeq
which is the quantum Liouville equation.

Two cases should be considered concerning the 
covariances $\rho_{jk}$ at. If at the initial
time, $\rho_{jk}=z_j^*z_k$, this form will be
preserved at all times and $z_j$ is given by
the equation 
\beq
i\mu \frac{dz_k}{dt} = \sum_j z_j H_{jk}, 
\label{61} 
\eeq
which is identified with equation (\ref{9})
and thus equivalent to the Schrödinger equation.
In this case ${\rm Tr} \rho^2 =({\rm Tr}\rho)^2 = 1$,
which corresponds to the quantum mechanics of pure
states. It should be pointed out that (\ref{61})
is a consequence of the quantum Liouville equation
(\ref{62}). In other words, $\rho_{jk}=z_j^*z_k$
solves the equation (\ref{62}) as long as $z_j$
satisfies the equation (\ref{61}). We remark
in addition that $z_j$ is not the average
$\la x_j\ra$. In fact $\la x_j\ra$ vanishes
for long times whereas $z_j$ does not in general
because
\beq
\sum_j z_j^*z_j = 1,
\eeq
which follows from (\ref{33a}).
If ${\rm Tr} \rho^2<1$, then it is not possible to 
write $\rho_{jk}$ as a product $z_j^*z_k$, and
this corresponds to the quantum mechanics of mixed
states. 

The average $\bar{A}$ of a state function
\beq
{\cal A} = \sum_{jk} A_{jk}x_j^* x_k
\eeq
is given by
\beq
\bar{A} =  \sum_{jk} A_{jk} \rho_{kj}.
= {\rm Tr} A\rho
\eeq
In the case of pure state,  $\rho=z z^\dagger$, 
where $z$ is a column matrix with elements $z_j$ and
$z^\dagger$ is the row matrix with elements $z_j^*$,
and $\bar{A}$ is reduced to the usual
quantum average $\bar{A} = z^\dagger A z$. 

We have considered above a noise that could change the
phase of the variable $x_k$ but not its absolute value.
This lead us to the the quantum Liouville equation and
to the Schrödinger equation. We consider now a more
generic noise that allows us to reach the Lindblad
equation that describes open quantum systems
\cite{lindblad1976,breuer2002}.

We add a white noise to equation (\ref{9})
which now reads
\beq
\frac{dx_j}{dt} = f_j + \zeta_j,
\label{9a}
\eeq
where
\beq
f_j = \frac1{i\mu}\sum_k H_{jk}x_k,
\eeq
and $\zeta_j$ is a stochastic variables with zero mean
that we choose to be linear in the variables $x_j$.
The noise should be chosen to conserve in the strict
sense the norm (\ref{33}) along a stochastic trajectory.
However, we relax this condition and requires that it
is conserved on the average.

A precise meaning of the stochastic equation (\ref{9a})
is provided by writing it in a  discrete time version
\cite{tome2022}, which is
\beq
\Delta x_j = \tau f_j + i\sqrt{\tau} \sum_k G_{jk}x_k
- \frac\tau2 \sum_k K_{jk} x_k
\label{14}
\eeq
where $\tau$ is the time interval and $\Delta x_j$
is the corresponding increment in the dynamical
variable $x_j$. The quantities $G_{jk}$ and
$K_{jk}$ are random variables to be found in such
a way that the norm (\ref{33}) is preserved.

Let us determine the increment in $x_jx_k^*$
during an interval of time $\tau$,
\beq
\Delta (x_jx_k^*) = x_j\Delta x_k^*
+ x_k^*\Delta x_j +\Delta x_j\Delta x_k^*.
\eeq
Using (\ref{14}), we find up to terms of order $\tau$
\[
\Delta (x_j x_k^*) = x_j \tau f_k^* +x_k^* \tau f_j
\]
\[
+ i\sqrt{\tau} \sum_n
(G_{jn}x_n x_k^* - G_{kn}^* x_j x_n^*)
+\tau\sum_{n\ell} G_{kn}^* G_{j\ell}x_\ell x_n^*
\]
\beq
- \frac\tau2 \sum_n (K_{kn}^* x_j x_n^* + K_{jn} x_n x_k^*).
\label{47}
\eeq
If we let $k=j$ in this equation and sum in $j$, we
find the increment in the norm (\ref{33}), which is
\[
\Delta{\cal N} = 
i\sqrt{\tau} \sum_{jn}(G_{nj} - G_{jn}^* )x_j x_n^*
\]
\beq
+\frac\tau2 \sum_{n\ell} (2\sum_j G_{jn}^* G_{j\ell}
- K_{\ell n}^*- K_{n\ell}) x_\ell x_n^*.
\eeq
We choose $K_{jk}$ so that the second summation vanishes,
that is,
\beq
K_{jk} = \sum_\ell G_{\ell j}^* G_{\ell k}.
\eeq

Next we choose $G_{jk}=g_{jk}\xi_{jk}$,
where $\xi_{jk}$ are real stochastic variables 
with zero mean and covariances $\la\xi_{jk}\xi_{\ell n}\ra=1$.
If we require $\Delta{\cal N}$ to vanish
in the strict sense, that is, in any stochastic
trajectory then $g_{jk}$ should equal $g_{kj}^*$,
resulting in the vanishing of the first summation
of (\ref{47}).
However, we require $\Delta{\cal N}$
to vanish in the average so that no restriction in
$g_{jk}$ is needed as the first summation of
(\ref{47}) will vanish in the average.

Taking the average of both sides of equation (\ref{47}),
the terms proportional to $\sqrt{\tau}$ vanish,
resulting in the following expression for
the time evolution of $\rho_{jk}=\la x_jx_k^*\ra$,
\[
\frac{d\rho_{jk}}{dt} = \frac1{i\mu}\sum_\ell
(H_{j\ell}\rho_{\ell k} - \rho_{j\ell} H_{\ell k})
+ \sum_{n\ell}  g_{j\ell}\rho_{\ell n}g_{kn}^*
\]
\beq
- \frac12 \sum_{n\ell} (\rho_{jn} g_{\ell n}^*g_{\ell k}
+ g_{\ell j}^* g_{\ell n}\rho_{nk}).
\eeq
Denoting by $g$ the matrix with elements $g_{jk}$,
this equation can be written in the form
\beq
\frac{d\rho}{dt} = \frac1{i\mu}[H,\rho] + \frac12
(2g\rho g^\dagger -\rho g^\dagger g - g^\dagger g \rho) 
\eeq
which is the Lindblad equation for open quantum systems
\cite{lindblad1976,breuer2002}.

We summarize our findings as follows. The dynamics
of a quantum system was shown to be represented by
an underlying classical system, which turns out
to be a collection of interacting classical harmonic
oscillators. The coordinate and momentum of the
classical particles are understood as the real
an imaginary parts of the wave function. 
The probabilistic character of quantum mechanics
is introduced explicitly by treating the wave
function as a time dependent random variable
by adding a white noise to the Hamilton equations
of motion that preserves the norm of the wave
function. The Schrödinger equation and the 
quantum Liouville equations are obtained when 
the noise changes the phase but not the absolute
value of the wave function. 

The present representation obviously does not transform
the science of quantum mechanics into the science of
classical mechanics. The underlying classical system
is not an observable as much as the wave function is
not. However, the present representation allows 
an interpretation of quantum mechanics other than
the standard interpretation
\cite{omnes1994,auletta2001,freire2022}.
As the trajectory in the Hilbert space is stochastic
this representation fits the consistent history
interpretation of quantum mechanics \cite{griffiths2002}
if we bear in mind that each possible trajectory is 
a possible history.


\end{document}